\documentclass[12pt,preprint]{aastex}

\def\kms{\ifmmode{\rm km\thinspace s^{-1}}\else km\thinspace s$^{-1}$\fi}

\begin{document}

\title{T-Lyr1-17236: A Long-Period Low-Mass Eclipsing Binary}

\author{Jonathan Devor\altaffilmark{1,2}, David Charbonneau\altaffilmark{1,3}, Guillermo Torres\altaffilmark{1},\\
Cullen H. Blake\altaffilmark{1}, Russel J. White\altaffilmark{4}, Markus Rabus\altaffilmark{5}, Francis T. O'Donovan\altaffilmark{6},\\
Georgi Mandushev\altaffilmark{7}, Gaspar Bakos\altaffilmark{1}, G{\'a}bor F\H{u}r{\'e}sz\altaffilmark{1}, and Andrew Szentgyorgyi\altaffilmark{1}}

\altaffiltext{1}{Harvard-Smithsonian Center for Astrophysics, 60 Garden Street, Cambridge, MA 02138}
\altaffiltext{2}{Email: jdevor@cfa.harvard.edu}
\altaffiltext{3}{Alfred P. Sloan Research Fellow}
\altaffiltext{4}{Physics Department, University of Alabama in Huntsville, Huntsville, AL 35899}
\altaffiltext{5}{Instituto de Astrof\'{\i}sica de Canarias, La Laguna, Tenerife, Spain}
\altaffiltext{6}{California Institute of Technology, 1200 East California Boulevard, Pasadena, CA 91125}
\altaffiltext{7}{Lowell Observatory, 1400 West Mars Hill Road, Flagstaff, AZ 86001}

\begin{abstract}

We describe the discovery of a 0.68+0.52~$M_{\sun}$ eclipsing
binary (EB) with an 8.4-day orbital period, found through a
systematic search of ten fields of the Trans-atlantic Exoplanet
Survey (TrES). Such long-period low-mass EBs constitute critical
test cases for resolving the long standing discrepancy between the
theoretical and observational mass-radius relations at the bottom
of the main sequence. It has been suggested that this discrepancy
may be related to strong stellar magnetic fields, which are not
properly accounted for in current theoretical models. All
previously well-characterized low-mass main sequence EBs have
periods of a few days or less, and their components are therefore
expected to be rotating rapidly as a result of tidal
synchronization, thus generating strong magnetic fields. In
contrast, the binary system described here has a period that is
over three times longer than previously characterized low-mass
main sequence EBs, and its components rotate relatively slowly. It
is therefore expected to have a weaker magnetic field and to
better match the assumptions of theoretical stellar models. Our
follow-up observations of this EB yield preliminary stellar
properties that suggest it is indeed consistent with current
models. If further observations confirm a low level of activity in
this system, these determinations would provide support for the
hypothesis that the mass-radius discrepancy is at least partly due
to magnetic activity.

\end{abstract}


\section{Introduction}
\label{secIntro}

\begin{figure}
\plotone{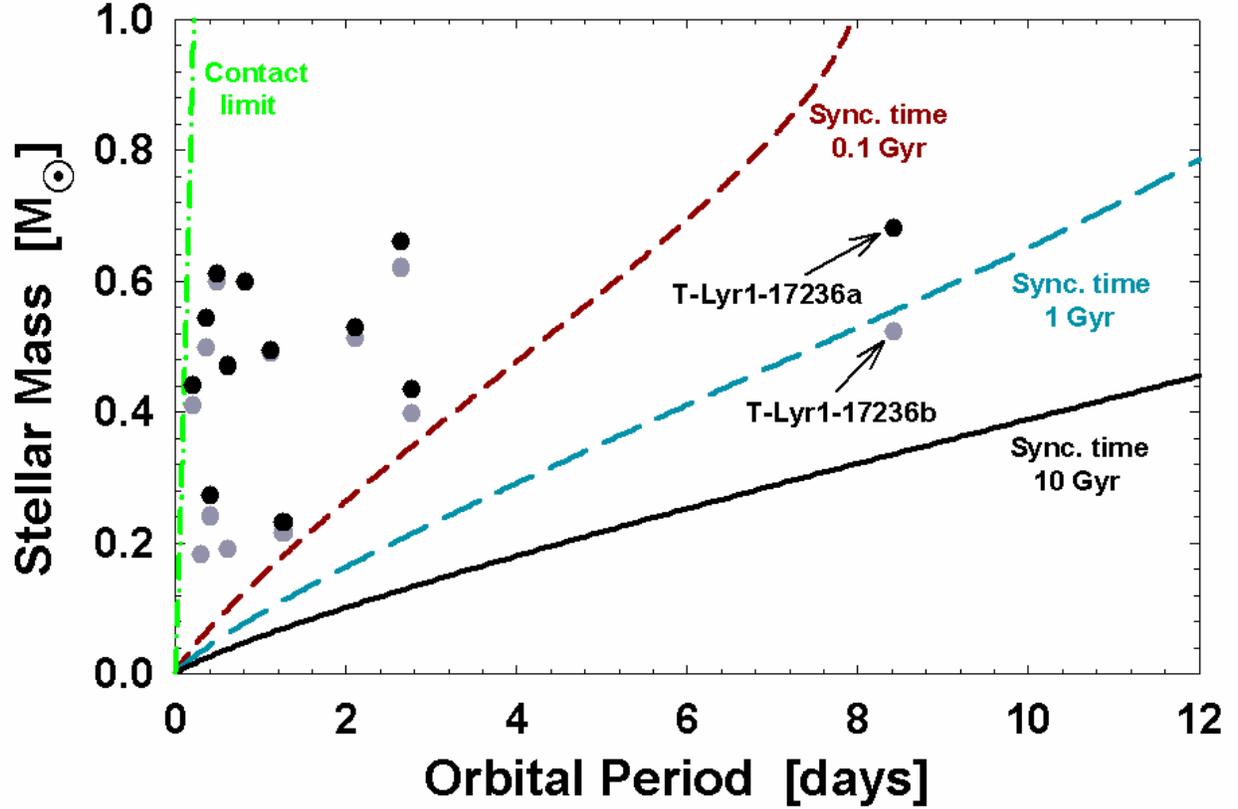}
\caption{The predicted synchronization timescales
due to turbulent dissipation \citep{Zahn77, Zahn94} for well
characterized low-mass EBs from Table~\ref{tablePreviousEBs}. The
lines trace constant synchronization timescales of binary
components for which $q=1$ (see \S\ref{secPhysical} for further
details on this calculation). The black circles indicate primary
components and the grey circles indicate secondary components.
Note that in some cases the primary and secondary symbols nearly
overlap.}
\label{figTsync}
\end{figure}

Despite a great deal of work that has been done to understand the
structure of low-mass ($<$ 0.8 $M_{\sun}$) main sequence stars
\citep[e.g.,][]{Chabrier00}, models continue to underestimate
their radii by as much as 15\% \citep{Lacy77rad, Torres02,
Ribas06}.  This is a significant discrepancy, considering that for
solar-type stars the agreement with the observations is typically
within 1--2\% \citep{Andersen91, Andersen98}. In recent years an
intriguing hypothesis has been put forward, suggesting that strong
magnetic fields may have bloated these stars, either through
chromospheric activity \citep[e.g.,][]{Ribas06, Torres06,
LopezMorales07, Chabrier07} or through magnetically induced
convective disruption \citep{Torres06}. Such strong magnetic
fields are expected to be formed by the dynamo mechanism of
rapidly rotating stars.\footnote{Dynamo theory predicts that this
mechanism operates only in partially convective stars. However,
the strong magnetic activity observed in fully convective low-mass
stars indicates that they also possess a mechanism for generating
strong magnetic fields \citep[see][and references
therein]{Browning07}.} To test this hypothesis, one needs to
measure both the masses and radii of low-mass stars, which thus
far can be done most accurately with eclipsing binary (EB)
systems. However, all well characterized low-mass main sequence
EBs have orbital periods shorter than three days (see
Table~\ref{tablePreviousEBs}) and are therefore expected to have
synchronization timescales shorter than $\sim$100 Myr \citep[][see
Figure~\ref{figTsync} and further description in
\S\,\ref{secPhysical}]{Zahn77, Zahn94}. As a result of these short
periods and short synchronization timescales, the rotations of
these binary components are expected to have accelerated to the
point that they now match the rapid angular velocity of their
orbits. With such rapid rotations, these binary components could
have a wide range of dynamo-induced magnetic field strengths. To
better constrain current stellar models, we set out to find
systems with slowly rotating components. Such systems would
presumably have comparably weak magnetic fields, thus being more
consistent with the model assumptions. Furthermore, by comparing
the mass-radius relations of binary components with well
determined levels of magnetic activity, one could test various
magnetic disruption models.

We note here that in addition to EB analysis, long-baseline optical
interferometry has also been used recently to measure the radii of
nearby low-mass stars \citep{Lane01, Segransan03, Berger06}.  While
these stars are single and are therefore expected to rotate slowly,
their masses can only be estimated through empirical mass-luminosity
relations or other indirect methods. Those determinations are thus
less fundamental, in a sense, and arguably of lesser value for
accurately constraining stellar models and testing the magnetic
disruption hypothesis.

\placefigure{figTsync}

\section{Initial Photometric Observations}
\label{secInitial}

T-Lyr1-17236 was first identified as a likely low-mass EB
candidate in the \cite{Devor08} catalog, following a systematic
analysis of the light curves (LCs) within ten fields of the
Trans-atlantic Exoplanet Survey \citep[TrES;][]{Alonso04}. TrES
employs a network of three automated telescopes to survey
$6\arcdeg \times 6\arcdeg$ fields of view. To avoid potential
systematic noise, we performed our initial search using data from
only one telescope, Sleuth, located at the Palomar Observatory in
Southern California \citep{ODonovan04}, and we combined additional
data at subsequent follow-up stages. Sleuth has a 10-cm physical
aperture and a photometric aperture radius of 30\arcsec. The
number of LCs in each field ranges from 10,405 to 26,495, for a
total of 185,445 LCs.  The LCs consist of $\sim$2000 Sloan
$r$-band photometric measurements binned to a 9-minute cadence.
The calibration of the TrES images, identification of stars
therein, and the extraction and decorrelation of the LCs are
described elsewhere \citep{Dunham04, Mandushev05, ODonovan06,
ODonovan07}.

An automated pipeline was used to identify and characterize the
EBs among the TrES LCs. This pipeline has been described in detail
in a previous paper \citep{Devor08}. At the heart of this analysis
lie two computational tools: the Detached Eclipsing Binary Light
curve fitter\footnote{The DEBiL source code, utilities, and
example files are available online at:\newline
http://www.cfa.harvard.edu/$\sim$jdevor/DEBiL.html}
\citep[DEBiL;][] {Devor05}, and the Method for Eclipsing Component
Identification\footnote{The MECI source code and running examples
are available online at:\newline
http://www.cfa.harvard.edu/$\sim$jdevor/MECI.html}
\citep[MECI;][]{Devor06a, Devor06b}. DEBiL fits each LC to a
$\it{geometric}$ model of a detached EB that consists of two
luminous, limb-darkened spheres that describe a Newtonian two-body
orbit. MECI then incorporated some of the DEBiL results, and
together with 2MASS color information \citep{Skrutskie06}, refit
each LC to a $\it{physical}$ model that is constrained by the
solar metallicity Yonsei-Yale theoretical isochrones \citep{Yi01,
Kim02}. Thus, using only photometric data, the DEBiL/MECI pipeline
provided initial estimates of the absolute physical properties of
each EB. These estimates were then used to locate promising
candidates for follow-up.

Using this pipeline a total of 773 EBs were identified within the
TrES dataset. Of these, 427 EBs were both detached and had small
out-of-eclipse distortions, thereby enabling the DEBiL/MECI
pipeline to estimate their component masses. These results,
together with many other properties, are listed for each EB in an
online
catalog\footnote{http://www.cfa.harvard.edu/$\sim$jdevor/Catalog.html}
\citep{Devor08}. Of these characterized EBs, we then identified a
handful of promising long-period low-mass candidates and chose
one, T-Lyr1-17236 ($\alpha_{2000}=19^h07^m16^s.621,\
\delta_{2000}=+46\arcdeg39'53''.21,\ P=8.429441 \pm 0.000033$
days\,; see Table~\ref{tableObjectParam} for additional
information), for further follow-up and analysis. As with all of
our low-mass candidates, we repeated the MECI analysis using the
\cite{Baraffe98} solar-metallicity isochrones (with a mixing
length parameter of $\alpha_{\rm ML} = 1.0$), which are more
accurate than the Yonsei-Yale isochrones in this regime. The
resulting MECI mass-mass likelihood contour plot of T-Lyr1-17236
is shown in Figure~\ref{figMECI}.  Since the MECI analysis
incorporates data from theoretical stellar models, we cannot use
it to constrain stellar models. Rather, once we identified the
candidate, we followed it up photometrically and
spectroscopically, and used only these follow-up data to derive
the binary's absolute properties.

\placefigure{figMECI}

\section{Follow-up Photometric Observations}

In order to characterize T-Lyr1-17236 we combined photometric data
from four telescopes: (1) Sleuth and (2) PSST \citep{Dunham04} of
the TrES network, (3) the Instituto de Astrof\'{\i}sica de
Canarias telescope \citep[IAC80;][]{Galan87}, and (4) the
Hungarian Automated Telescope Network \citep[HATNet;][]{Bakos04}.
With the exception of the IAC80, we obtained our photometric data
from archived survey datasets that were intended for locating
exoplanets.

As part of the TrES network (see \S\,\ref{secInitial}), Sleuth and
PSST are operated similarly. However, PSST, which is located at
the Lowell Observatory in Arizona, observes in the Johnson
$R$-band whereas Sleuth observes in the Sloan $r$-band (see
Figures~\ref{figLC_lyr17236} and \ref{figEclipse_r}). Furthermore,
PSST has a 20\arcsec\ photometric aperture radius compared to
Sleuth's 30\arcsec\ radius, which provides PSST with a higher
resolving power than Sleuth. However, the smaller aperture of PSST
also causes it to have noisier photometry, with an RMS of 0.031
mag for T-Lyr1-17236, compared to the Sleuth photometry that has
an RMS of 0.028 mag. Though these differences are small, they
would have affected our analysis. We therefore chose not to use
the PSST data for fitting the photometric model, though we did use
them to improve the determination of the orbital period and the
epoch of eclipse (see \S\,\ref{secAnalysis}).

In an effort to better constrain the eclipses of T-Lyr1-17236, we
obtained data from the IAC80, an 82-cm aperture telescope with a
$14' \times 14'$ field of view, located at the Observatorio del
Teide in the Canary Islands. We produced an $I$-band LC at a
1.3-minute cadence using the 1024$\times$1024-pixel Tromso CCD
Photometer (TCP), resulting in 0.008 mag RMS photometry for
T-Lyr1-17236. Unfortunately, we were only able to observe a
primary eclipse with the IAC80. We therefore incorporated archival
HATNet observations so as to provide coverage of the secondary
eclipse in a similar bandpass (see Figures~\ref{figLC_lyr17236}
and \ref{figEclipse_I}).

HATNet is a network of six 11-cm aperture, fully-automated
telescopes (HATs) located at the F. L. Whipple Observatory in
Arizona and at the Submillimeter Array site atop Mauna Kea,
Hawaii. The HATs have an $8\arcdeg \times 8\arcdeg$ field of view,
a response that peaks in the $I$-band, and operate at a 5.5-minute
cadence. To reduce the photometric noise, the HAT point spread
function (PSF) is broadened to a $\sim$15\arcsec\ aperture radius
through microstepping \citep{Bakos02}. Even so, the HATNet
photometric RMS for T-Lyr1-17236 was comparably large, at 0.084
mag. Nevertheless, to provide more complete coverage of the
primary and secondary eclipses in the $I$-band, we combined the
IAC80 observations with data from HAT-7 (Whipple Observatory) and
from HAT-8 (Mauna Kea). Due to the very different characteristics
of these two systems, however, we chose not to adopt any of the
model parameters derived from these data, and only used these
results as an independent confirmation of the Sleuth $r$-band LC
analysis.

\placefigure{figLC_lyr17236}
\placefigure{figEclipse_r}
\placefigure{figEclipse_I}

\section{Spectroscopic Observations}
\label{secSpecObs}

T-Lyr1-17236 was observed spectroscopically with two instruments: The
Near-Infrared Spectrometer \citep[NIRSPEC;][]{McLean98, McLean00} at
the W.~M.~Keck Observatory in Hawaii, and the Tillinghast Reflector
Echelle Spectrograph \citep[TRES;][]{Szentgyorgyi07}, installed on the
1.5-meter Tillinghast telescope at the F.\ L.\ Whipple Observatory in
Arizona.

NIRSPEC was operated using a 3-pixel slit (0.432\arcsec) and an N7
blocking filter, thus producing a spectral resolving power of
\textsf{R} $= \lambda/\Delta\lambda \simeq 25,\!000$. The duration
of the exposures, which ranged from 420 to 900 seconds, was
adjusted according to observing conditions. The spectra were
gathered in two consecutive nods, producing a total of five
NIRSPEC nod pairs.  The nods of each pair were then subtracted one
from the other, removing much of the sky emission. We extracted
the spectra of both nods using the optimal extraction procedure
outlined in \citet{Horne86}, and then co-added the two resulting
one-dimensional spectra. We calibrated the wavelengths of the
resulting spectrum using its atmospheric telluric features, and
then corrected for both the telluric absorption and the blaze of
the spectrograph by dividing this spectrum by the spectrum of an
A0V-type star (HR~5511). Finally, we cross-correlated each
spectrum with the spectrum of an M0.5V template star (GJ~182). To
this end, we used a single NIRSPEC order (2290--2320~nm), which is
within the $K$-band, and has a scale of $0.0336\:$nm\,pixel$^{-1}$
at its center. This order covers the CO~2-0 bandhead, which
includes a rich forest of R-branch transition lines, as well as
many telluric absorption features due to methane in the Earth's
atmosphere. The advantages offered by this spectral region and the
details of the instrument setup are described in \citet{Blake08}.

TRES is a high-resolution fiber-fed optical echelle spectrograph
designed to cover a large range of wavelengths (390--934~nm) in 51
orders. We employed the medium-size fiber (2.3\arcsec) so as to
cover the full stellar PSF, while providing a spectral resolving
power of \textsf{R} $\simeq 47,\!000$. Following each of our three
900--1000 second exposures, the TRES data were read from a
4638$\times$1090-pixel CCD, which we set to a 2$\times$2 binning
mode for a more rapid read-out. We then used a dedicated IRAF
toolset to process and extract 51 spectral orders simultaneously,
ultimately producing 2319 data points along each order. The IRAF
processing of the TRES data involved merging the mosaic FITS
files, removing cosmic ray hits, flattening fringing effects, and
then extracting the orders. We wavelength-calibrated the TRES
spectra using Thorium-Argon (ThAr) exposures, and then corrected
the telluric absorption and spectroscopic blazing by dividing each
spectrum by a TRES spectrum of a rapidly-rotating B0IV-type star
(HR~264). Though TRES produces 51 spectral orders, we used only
four of them, covering wavelengths of 665--720~nm (similar to the
$R$-band), and having a post-binning scale of
$\sim$0.0065$\:$nm\,pixel$^{-1}$. These orders contain a diverse
array of absorption features, including those of TiO, Fe$\:$I,
Ca$\:$I, Ni$\:$I, and Cr$\:$I. We limited ourselves to these
orders because at shorter wavelengths there was insufficient flux
from our red target, while at longer wavelengths the spectra were
dominated by telluric absorption features, produced largely by
terrestrial O$_2$ and H$_2$O. We cross-correlated these four
orders with the corresponding orders of an M1.5V template star
(GJ~15A, also known as GX~And~A) and averaged their
cross-correlation functions. We repeated this final calculation
using the \citet{Zucker03} maximum-likelihood method, which
reproduced our results to within a fraction of their
uncertainties, although with slightly larger errors.\footnote{The
\citet{Zucker03} method is more accurate than simple
cross-correlation averaging for large N. However, because it takes
the absolute value of the correlation, it loses some information
and effectively increases the noise baseline. This increased noise
will negate its advantage when combining a small number of
correlations, as is the case in our TRES analysis ($N = 4$).}

In total, we produced five RV measurements of each component with
NIRSPEC and three with TRES. In all cases we were able to measure
the RVs of both binary components by employing a cross-correlation
method that transforms the spectra to Fourier-space using the
Lomb-Scargle algorithm \citep{Press92}.  This method allowed us to
cross-correlate spectra with arbitrary sampling, without having to
interpolate or resample them onto an equidistant grid. We then
multiplied the Fourier-transformed target and template spectra,
inverse-Fourier-transformed the product, and normalized it. Since
the resulting two peaks in the cross-correlation functions were
always well separated, we were able to fit each with a parabola,
and thus measure their offsets and widths. The uncertainties of
these RVs are somewhat difficult to determine with our procedures,
but tests indicate that they are approximately 1.0~\kms\ and
1.4~\kms\ for the primary and secondary in our NIRSPEC spectra,
and about 0.5~\kms\ and 1.2~\kms\ in our TRES spectra. These
internal errors are adopted below in the spectroscopic analysis,
but have relatively little effect on the results. Finally, the RVs
were transformed to the barycentric frame, and the TRES RV
measurements were further offset by $-$2.82~\kms\ in order to
place them on the same reference frame as the NIRSPEC
measurements, which were obtained with a different template
(GJ~182). This offset was determined by including it as an
additional free parameter in the Keplerian RV model (see
\S\,\ref{secAnalysis}). Once the offset was determined, we held
its value fixed in all subsequent analyses. The final velocities
are listed in Table~\ref{tableLyr17236RVs} and include this
offset. Note that these listed RVs are all relative to GJ~182, for
which \cite{Montes01} have measured the value $+32.4 \pm
1.0$~\kms.

\section{Orbital Analysis}
\label{secAnalysis}

We began our analysis by determining the orbital period ($P$) and
the epoch of primary eclipse ($t_0$), and constraining the
eccentricity ($e$) of T-Lyr1-17236 through eclipse timing. The
times of eclipse determined from our photometric observations
listed in Table~\ref{tableLyr17236Timing}. Since our data span 3.5
years, we were able to determine the period to an accuracy of 3
seconds (see Table~\ref{tablePhotoParam}). To estimate the
binary's eccentricity, we first measured the observed minus
calculated ($O\!-\!C$) timing difference between the primary and
secondary eclipses in all available LCs, which provided an upper
bound of $|e\cos\omega| \lesssim 0.0008$, where $\omega$ is the
argument of periastron (see Figure~\ref{figTiming}). Though
$\omega$ and $e$ cannot be determined separately in this way, this
result indicates that the orbit of T-Lyr1-17236 is likely to be
circular or very nearly so. This conclusion is further supported
by a weaker upper limit of $|e\sin\omega| \lesssim 0.06$, obtained
through preliminary LC model fitting (see below). Theoretical
estimates \citep{Zahn77, Zahn78, Zahn94} of this binary suggest a
circularization timescale of $t_{\rm circ} \simeq 390$~Gyr
\citep[see also][]{Devor08}. Being many times the age of the
binary, this long timescale suggests that T-Lyr1-17236 formed in a
circular orbit. However, this timescale value is an instantaneous
estimate for the current epoch, and is likely to have been
significantly different in the past \citep[see][and references
therein]{Zahn89, Mazeh08}. Therefore, it is quite possible that
the binary circularized while it was in the pre-main sequence,
however, to the extent that this theory is correct, it is unlikely
to have circularized once settling on the main sequence.

\placefigure{figTiming}

A Keplerian model was fitted to the radial velocities to determine
the elements of the spectroscopic orbit of T-Lyr1-17236. We
assumed the eccentricity to be zero based on the evidence above
and the lack of any indications to the contrary from preliminary
spectroscopic solutions. The period and $t_0$ were held fixed at
the values determined above. We solved simultaneously for the
velocity semi-amplitudes of the components ($K_{A,B}$) and the RV
of their center of mass ($V_\gamma$). The results are shown
graphically in Figure~\ref{figPhasedRV}, and the elements are
listed in Table~\ref{tableSpecParam}. The minimum masses $M_{A,B}
\sin^3 i$ are formally determined to better than 2\%. However,
because of the small number of observations ($N = 8$), the
possibility of systematic errors cannot be ruled out and further
observations are encouraged to confirm the accuracy of these
results.

\placefigure{figPhasedRV}

We then proceeded to find the remaining photometric parameters of
T-Lyr1-17236. To this end, we analyzed the Sleuth $r$-band LC using
JKT-EBOP \citep{Southworth04a, Southworth04b}, a LC modeling program
based on the EPOB light curve generator \citep{Nelson72, Etzel81,
Popper81}. We assumed a circular orbit, as before, a mass ratio of $q
= 0.7692$ from the spectroscopic model, and the period determined
above. We solved simultaneously for the orbital inclination ($i$), the
fractional radii ($r_{A,B}$), the central surface brightness ratio of
the secondary in units of the primary ($J$), the time of primary
eclipse ($t_0$), and the out-of-eclipse magnitude (zero point). We
estimated the uncertainties of the fitted parameters by evaluating the
distribution generated by 1000 Monte Carlo simulations
\citep{Southworth05}.

Because of the large photometric aperture of Sleuth, the presence
of significant contamination from the light of additional stars is
a distinct possibility. Unfortunately, due to its degeneracy with
the orbital inclination and the fractional radii, we were not able
to simultaneously determine the fractional third light of the
system ($l_3$). We therefore sequentially refit the LC model
parameters with fixed fractional third-light values ranging from 0
to 0.2 (see Figure~\ref{fig3light}). We repeated this routine with
the $I$-band IAC80/HATNet LC as well, although these results were
not used because of their larger uncertainties. We obtained an
external estimate of the third-light fraction affecting the Sleuth
observations using the USNO-B catalog \citep{Monet03}, which lists
two dim objects within 30\arcsec\ of T-Lyr1-17236 (USNO-B1.0
1366-0314297 and 1366-0314302). Assuming that these objects are
completely blended into T-Lyr1-17236, we expect an $R$-band
third-light fraction of $l_3=0.085 \pm 0.018$, and we adopted this
value for the $r$-band LC. Fortunately, the fitted parameters are
quite insensitive to third light, so that the uncertainty in $l_3$
only moderately increases their uncertainties.  No objects were
listed within the smaller photometric apertures of either IAC80 or
HATNet, so we conclude that the $I$-band LC should have little or
no third-light contamination. It is important to note that these
third-light estimates assume that there are no further unresolved
luminous objects that are blended with T-Lyr1-17236 (e.g., a
hierarchical tertiary component).  However, the divergence of the
$r$-band and $I$-band solutions at higher third-light fractions
(see Figure~\ref{fig3light}), and the deep primary eclipse in both
the $r$- and $I$-bands (0.649 mag and 0.604 mag, respectively),
suggest that if such unresolved objects exist, they are unlikely
to account for more than $\sim$0.1 of the total flux, and
therefore would not bias the fitted results beyond the current
estimated uncertainties. The final results of our LC fits are
given in Table~\ref{tablePhotoParam}.

\placefigure{fig3light}

\section{Physical Parameters}
\label{secPhysical}

The fundamental parameters of T-Lyr1-17236, such as their absolute
masses and radii, were derived by combining the results of the
spectroscopic analysis (Table~\ref{tableSpecParam}) with those
from the photometric analysis (Table~\ref{tablePhotoParam}). These
and other physical properties are listed in
Table~\ref{tableSystemParam}. Our estimates of the primary and
secondary component masses, $M_A$ = 0.6795 $\pm$ 0.0107 $M_{\sun}$
and $M_B$ = 0.5226 $\pm$ 0.0061 $M_{\sun}$, lead us to infer
spectral types of K5V and M0V, respectively, according to
empirical tables \citep{Cox00}. We are not able to make
independent estimates of the effective temperatures of the stars
from the data in hand. This could be done, for example, if we had
individual color indices based on combined light values and light
ratios in two different bands, but we can only derive a reliable
estimate of the light ratio in the $r$-band. The comparison with
stellar evolution models by \cite{Baraffe98} in
\S\,\ref{secConclusions} suggests primary and secondary component
temperatures of approximately 4150~K and 3700~K, respectively,
although the accuracy of these values is difficult to assess.

No trigonometric parallax is available for T-Lyr1-17236.  A rough
distance estimate to the system may be made using the $JHK_s$
brightness measurements in the 2MASS Catalog, collected in
Table~\ref{tableObjectParam}, along with estimates of the absolute
magnitudes. For these we must rely once again on models. The
Galactic latitude of $+16.8\arcdeg$ suggests the possibility of
some interstellar extinction. From the reddening maps of
\cite{Schlegel98} we infer $E(B-V) \simeq 0.07$ in the direction
of the object (total reddening), which corresponds to extinctions
of $A(J) \simeq 0.061$, $A(H) \simeq 0.038$, and $A(K) \simeq
0.011$, assuming $R_V = 3.1$ \citep{Cox00}. Under the further
assumption that this extinction applies to T-Lyr1-17236, we derive
a mean distance of $230 \pm 20$~pc, after conversion of the
near-infrared magnitudes in the CIT system from the
\cite{Baraffe98} models to the 2MASS system, following
\cite{Carpenter01}. With the proper motion components from the
USNO-B Catalog listed in Table~\ref{tableObjectParam}, the
center-of-mass velocity $V_\gamma$ from the spectroscopic solution
corrected for the velocity of GJ~182 \citep{Montes01}, and the
distance above, we infer space velocity components in the Galactic
frame of ($U$,$V$,$W$) $\simeq$ ($+41$,$+21$,$+2$)~\kms, where $U$
points in the direction of the Galactic center.

Because of the relevance of the rotational velocities of the stars
for the interpretation of the chromospheric activity results of
\S\,\ref{secActivity}, we have made an effort here to measure the
rotational broadening of both components from the widths of the
cross-correlation functions derived from our TRES spectra. We rely
on the fact that to first order, the width of a cross-correlation
peak is approximately equal to the quadrature sum of the line
broadening of the two spectra. We began our estimation procedure
by finding the effective resolution of the instrument ($\sigma_i$)
in the four TRES orders we used. This was done by auto-correlating
a TRES ThAr spectrum that was taken just before the second
T-Lyr1-17236 observation. We found that the four orders produced
peaks with an average FWHM of $8.90 \pm 0.17$~\kms.  Thus,
assuming that the intrinsic widths of the ThAr emission lines are
negligible compared to the instrumental resolution, we found that
$\sigma_i = 6.29 \pm 0.12$~\kms. This value corresponds to a
spectral resolving power of \textsf{R} $= 47,\!630 \pm 930$, which
is consistent with the TRES specifications. Next, we determined
the intrinsic spectral line broadening of the template star,
GJ~15A ($\sigma_t$). We auto-correlated the template spectrum and
found that it produced peaks with an average FWHM of $9.7 \pm
1.4$~\kms. This value should be equal to $\sqrt{2}(\sigma_i^2 +
\sigma_t^2)^{1/2}$, from which we infer that $\sigma_t = 2.7 \pm
2.5$~\kms. Note that this result is well within the upper bound
provided by \citet{Delfosse98}, following their non-detection of
any rotational broadening in GJ~15A. Using this information, we
can now find the intrinsic spectral line broadening of the
T-Lyr1-17236 components ($\sigma_{A,B}$). The average FWHM of the
primary and secondary peaks, resulting from the cross-correlation
of each observed spectrum of T-Lyr1-17236 against the template,
were measured to be $12.6 \pm 2.0$~\kms\ and $12.0 \pm 2.4$~\kms,
respectively. These widths are expected to be equal to
$[(\sigma_i^2 + \sigma_t^2) + (\sigma_i^2 +
\sigma_{A,B}^2)]^{1/2}$, from which we calculate that $\sigma_A =
8.4 \pm 3.0$~\kms\ and $\sigma_B = 7.6 \pm 3.8$~\kms.

The rotational profile FWHM expected for a homogeneous stellar
disk is $\sqrt{3}\,v \sin i_r$, where $v$ is the star's equatorial
rotational velocity, and $i_r$ is the inclination of its
rotational axis. Stellar limb darkening, however, will narrow the
rotational profile, thus decreasing the observed FWHM
\citep{Gray92}. Adopting the $R$-band PHOENIX linear limb
darkening coefficients from \cite{Claret98}, we find that the
expected FWHM values for the primary and secondary components of
T-Lyr1-17236 are, respectively, $1.495\,v \sin i_r$ and $1.499\,v
\sin i_r$. Using these results we can set upper bounds to the
components' $v \sin i_r$. These upper bounds represent the
limiting case whereby the spectral line broadening is due entirely
to stellar rotation, and we neglect all other line broadening
mechanisms, such as microturbulence and the Zeeman effect. We thus
determine the maximum rotational velocities of the T-Lyr1-17236
primary and secondary components to be $v \sin i_r = 5.6 \pm
2.0$~\kms\ and $5.1 \pm 2.3$~\kms, respectively.

An estimate of the timescale for tidal synchronization of the
stars' rotation with their orbital motion may be obtained from
theory following \cite{Zahn77}, and assuming simple power-law
mass-radius-luminosity relations \citep{Cox00}. Thus, for stars
less massive than 1.3~$M_{\sun}$,
\begin{equation}
\label{eqCircTime} t_{\rm sync} \simeq 0.00672~{\rm Myr} \; (k_2 / 0.005)^{-1}q^{-2}(1+q)^2 \left(P / {\rm day}\right)^4 \left(M / M_\sun \right)^{-4.82}~,
\end{equation}
where $k_2$ is determined by the structure and dynamics of the
star and can be obtained by interpolating published theoretical
tables \citep{Zahn94}. This calculation leads to timescales of
$t_{\rm sync} \simeq 0.56$~Gyr and 1.02~Gyr for the primary and
secondary components of T-Lyr1-17236, respectively, which are much
shorter than the circularization timescale determined in
\S\,\ref{secAnalysis}. We note that similar to the circularization
timescale, the synchronization timescales estimated above are the
current instantaneous values, and are likely to have changed over
time. The age of the system is undetermined (see
\S\,\ref{secConclusions}), but assuming its age is at least a few
Gyr, as is typical for field stars, it would not be surprising if
tidal forces between the components had already synchronized their
rotations. This is illustrated in Figure~\ref{figTsync}, where
T-Lyr1-17236 is shown along with the other systems in
Table~\ref{tablePreviousEBs} and with curves representing
theoretical estimates of the synchronization timescale as a
function of orbital period.

If we assume that the components are indeed rotationally
synchronized, we can compute their rotational velocities more
accurately using $v_{A,B} = 2\pi R_{A,B}/P$. We thus derive
synchronized velocities of $(v \sin i_r)_{\rm sync} = 3.81 \pm
0.26$~\kms\ and $3.15 \pm 0.31$~\kms\ for the primary and
secondary components, respectively. These values are slightly
below but still consistent with the maximum rotational velocities
measured above. Thus, observational evidence suggests that the
stars' rotations may well be synchronized with their orbital
motion, although more precise measurements would be needed to
confirm this. Our conclusion from this calculation is that
regardless of whether we assume that the components of
T-Lyr1-17236 are synchronized, their rotational velocities do not
appear to be large.

\section{Chromospheric Activity}
\label{secActivity}

Our absolute mass and radius determinations for T-Lyr1-17236 offer
the possibility of testing stellar evolution models in the lower
main sequence, and in particular testing the idea that the
discrepancies noted in \S\,\ref{secIntro} are related to
chromospheric activity and the associated magnetic fields in
systems where the components are rotating relatively rapidly.
Thus, establishing the level of the activity in the system
presented here is of considerable importance. We have shown in
\S\,\ref{secPhysical} that the relatively long period of
T-Lyr1-17236 ($P \simeq 8.429441$~days) implies that even if the
components are synchronized, their rotational velocities are slow,
and therefore are not expected to induce a great deal of
chromospheric activity. However, demonstrating that the stars are
indeed inactive requires more direct evidence, given that some
stars of similar masses as these are still found to be quite
active at rotation periods as long as 8 days \citep[see,
e.g.,][]{Pizzolato03}. We present here the constraints available
on the surface activity of T-Lyr1-17236 from its X-ray emission,
optical variations, and spectroscopic indicators.

The present system has no entry in the ROSAT Faint Source Catalog
\citep{Voges99}, suggesting the X-ray luminosity, usually
associated with activity, is not strong. Examination of the
original ROSAT archive images leads to a conservative upper limit
to the X-ray flux of $6.71 \times 10^{-14}$
erg$\thinspace$cm$^{-2}\thinspace$s$^{-1}$ in the energy range
0.1--2.4~keV, and together with information from
Table~\ref{tableSystemParam}, we infer an upper limit for the
ratio of the X-ray to bolometric luminosity of $\log L_X/L_{\rm
bol} \lesssim -3.13$\,. Values for the four best studied cases of
CM~Dra, YY~Gem, CU~Cnc, and GU~Boo, which are all very active, are
respectively $-3.15$, $-2.88$, $-3.02$, and $-2.90$
\citep[see][]{LopezMorales07}. These are at the level of our limit
or higher, although we do not consider this evidence conclusive.

There are no detectable variations in the $r$-band light curve out
of eclipse, within the uncertainties. Such variations would be
expected from activity-related surface features showing
significant contrast with the photospheres. We estimate an upper
limit of $\sim$0.01 mag in $r$ for the night-to-night variations
(see Figure 3). Because the secondary components is significantly
dimmer, it has a weaker variability upper limit of $\sim$0.09 mag.
We note, however, that this evidence for inactivity is not
conclusive either, since the observed photometric variations can
depend significantly on the distribution of spots on the surface.

A number of spectroscopic activity indicators (the Ca~II H and K
lines, H$\alpha$, etc.) should in principle allow a more direct
assessment of the activity level in T-Lyr1-17236.  Unfortunately,
however, the quality of our spectroscopic material in the optical
makes this difficult. The flux in the blue for this very red
system is too low to distinguish the Ca~II H and K lines, and even
at H$\alpha$ the noise is considerable (typical signal-to-noise
ratios at this wavelength are $\sim$12~pixel$^{-1}$). Two of the
three TRES spectra show the H$\alpha$ line in absorption, and the
other appears to show H$\alpha$ in emission. This suggests some
degree of chromospheric activity, although perhaps not at such a
high level as to sustain the emission at all times, as is seen in
other stars. H$\beta$ appears to be in absorption in all three
TRES spectra.

Clearly more spectra with higher signal-to-noise ratios are needed to
better characterize the level of activity, but from the sum of the
evidence above it would not appear that the activity in T-Lyr1-17236
is as high as in other low-mass eclipsing binaries studied previously,
thus more closely aligning it with the assumptions of current standard
stellar models. The system may therefore constitute a useful test case
for confirming or refuting the magnetic disruption hypothesis (see
\S\,\ref{secIntro}), which predicts that the absolute properties of
its slowly rotating components should match the theoretical models of
convective stars.

\section{Comparison with Models and Conclusions}
\label{secConclusions}

A comparison with solar-metallicity models by \cite{Baraffe98} for
a mixing length parameter of $\alpha_{\rm ML} = 1.0$ is presented
in Figure~\ref{figIsochrones}. Our mass and radius determinations
for T-Lyr1-17236 (see Table~\ref{tableSystemParam}) are shown
along with those of the low-mass systems listed in
Table~\ref{tablePreviousEBs}. The location of the models in this
diagram depends only slightly on age because these stars evolve
very slowly. The age of T-Lyr1-17236 is difficult to establish
independently. The space motions derived in \S\,\ref{secPhysical}
do not associate the system with any known moving group, and are
quite typical of the thin disk. Thus, all we can say is that it is
not likely to be very old. We display in
Figure~\ref{figIsochrones} two models for ages of $1$~Gyr and
$10$~Gyr, which likely bracket the true age of T-Lyr1-17236.
Within the errors, our measurements for the two components are
consistent with the models, which would in principle support the
magnetic disruption hypothesis. Unfortunately, however, the
uncertainties in the radius measurements ($\sim$7\% and
$\sim$10\%) are still large enough that our statement cannot be
made more conclusive. Further follow-up observations, especially
rapid-cadence and precise photometric measurements during multiple
eclipses, should significantly reduce the uncertainties in the
radii and thus provide far stronger constraints on the theoretical
models of low-mass stars. Additionally, higher-quality
spectroscopic observations than ours are needed to confirm that
the level of chromospheric activity in the system is relatively
low.  If after such observations, the masses and radii of the
T-Lyr1-17236 components remain consistent with the stellar models,
then the magnetic disruption hypothesis will be strengthened.
However, if further observations find that the components of
T-Lyr1-17236 are larger than predicted by current stellar models,
as is the case with most other similar systems investigated in
sufficient detail, then this will provide evidence that additional
mechanisms need to be included in the models of the structure of
low-mass main sequence stars \citep[see, e.g.,][]{Chabrier07}.

\placefigure{figIsochrones}

It is important to note here that T-Lyr1-17236 falls within the
field of view of the upcoming NASA Kepler Mission
\citep{Borucki03}. The Kepler Mission will not return data for all
stars within its field of view; rather, the targets will be
selected by the Kepler team. We see at least two reasons why such
monitoring of T-Lyr1-17236 would be of significant value. First,
the data would greatly refine the estimates of the physical
parameters of the component stars and may permit a search for
their asteroseismological modes. Second, the data would enable a
search for transits of exoplanets, which are expected to orbit in
the same plane as that defined by the stellar orbits.

\begin{figure}
\plotone{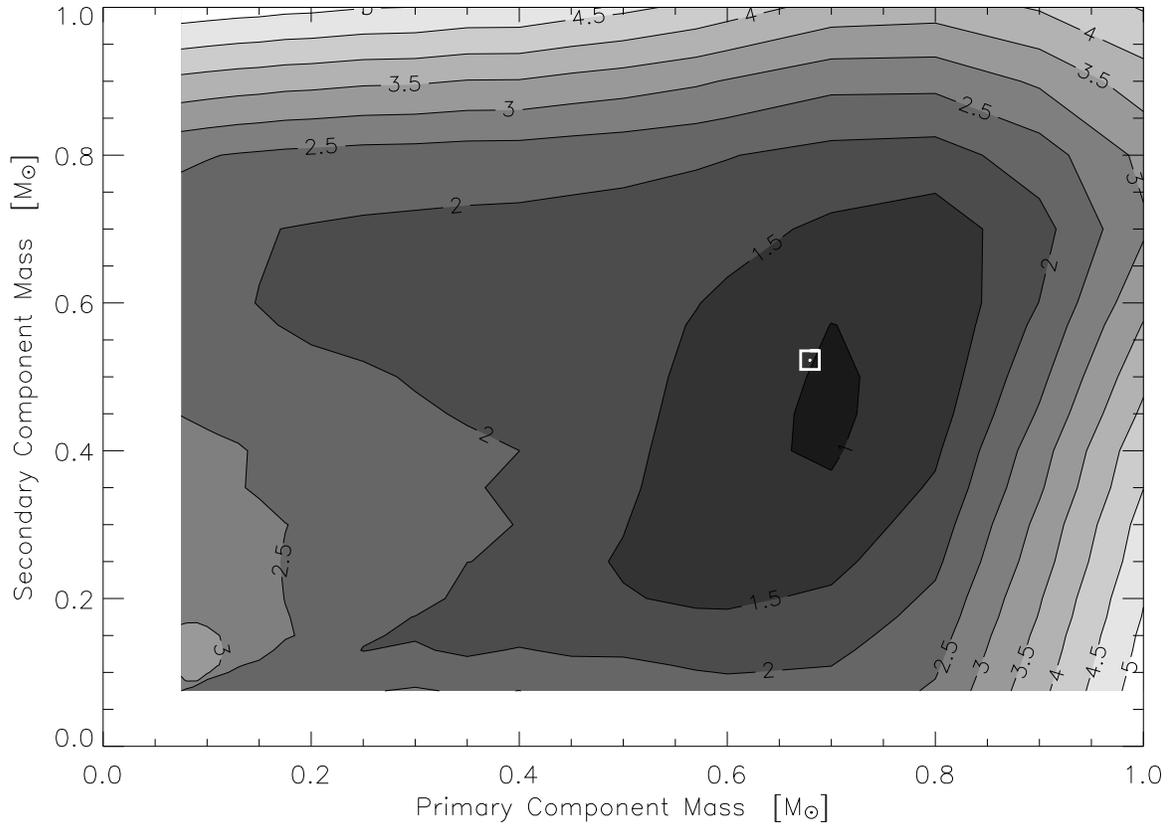}
\caption{The mass-mass likelihood plot for
T-Lyr1-17236 created with MECI, using the \cite{Baraffe98}
isochrones for an age of 2.5~Gyr. This analysis incorporated the
$r$-band LC and the 2MASS colors of the target. The contour lines
indicate the weighted reduced chi-squared values of each component
mass pairing, using $w=10$ \citep{Devor06b}. The white point
indicates the our final mass estimate from this paper, and the white
square approximates our current mass uncertainties.}
\label{figMECI}
\end{figure}

\begin{figure}
\plotone{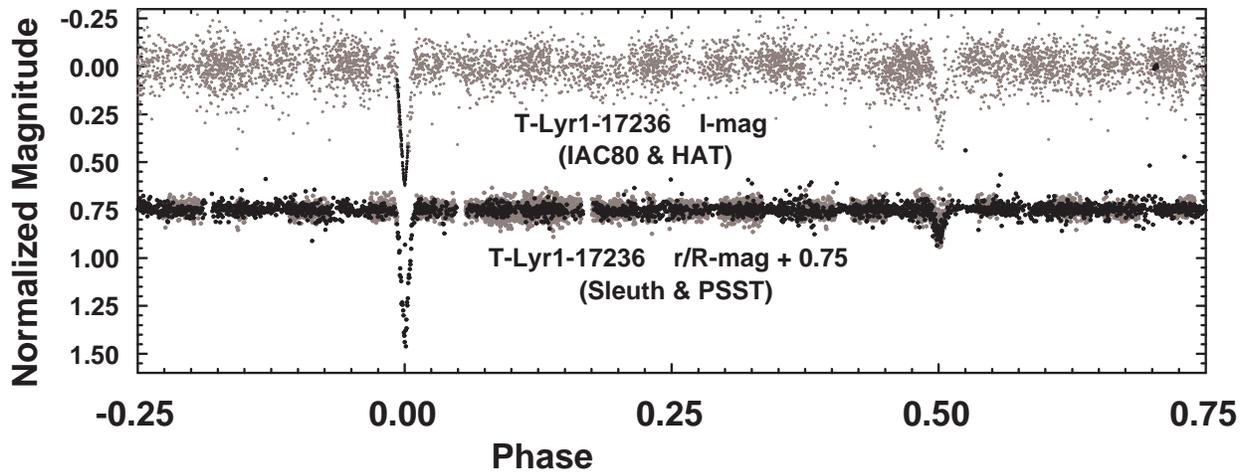}
\caption{Phased light curves of T-Lyr1-17236. The
top curve is from the IAC80 (black) and HATNet (grey) telescopes,
both of which observed in $I$-band. Note the tight cluster of
IAC80 observations near phase 0.7; these points determine the
IAC80 LC zero point. The bottom curve is from the Sleuth (black)
and PSST (grey) telescopes, which observe, respectively, in the
$r$-band and $R$-band. The secondary eclipse is about twice as
deep in the $I$-band as it is in the $r$- or $R$-bands, indicating
that the secondary component is significantly redder and
therefore cooler than the primary.}
\label{figLC_lyr17236}
\end{figure}

\begin{figure}
\plotone{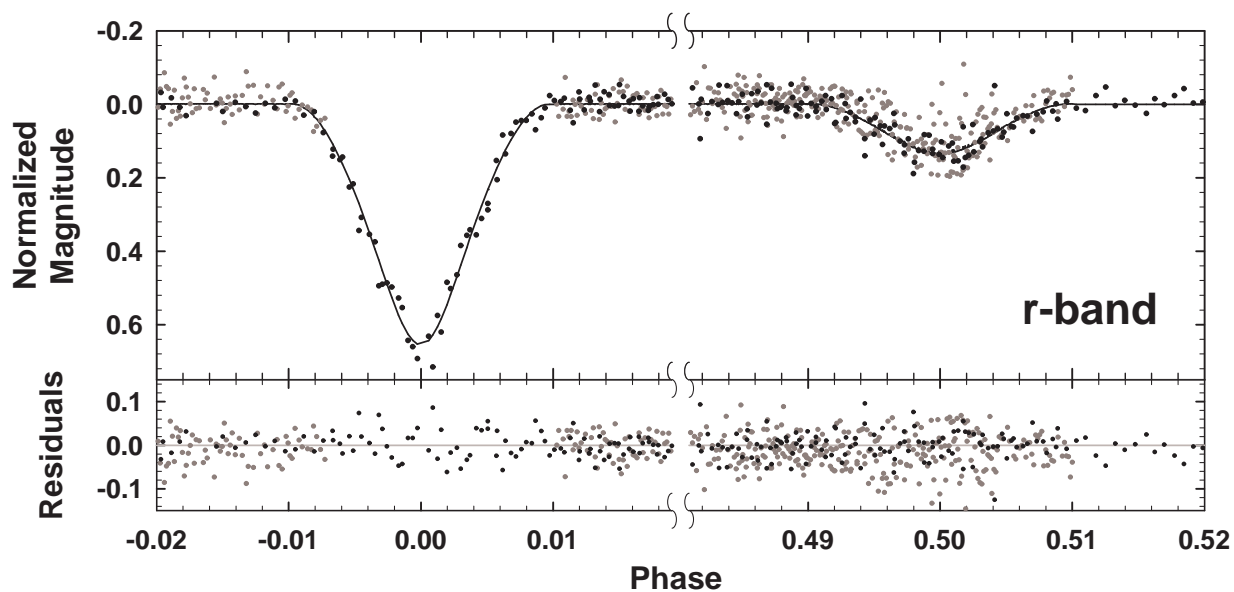}
\caption{Enlargement of the eclipse phases in the LC of T-Lyr1-17236,
as recorded by the Sleuth (black) and PSST (grey) telescopes ($r$-band
and $R$-band, respectively). The solid line shows the best-fit
JKT-EBOP model, for which the residuals are displayed at the bottom.}
\label{figEclipse_r}
\end{figure}

\begin{figure}
\plotone{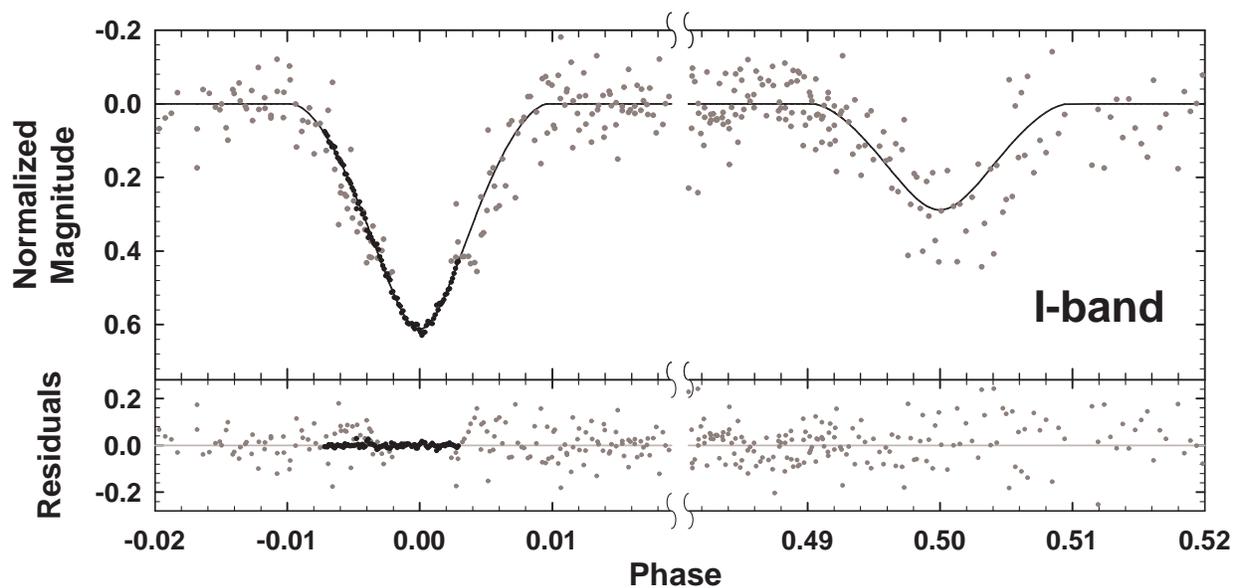}
\caption{Enlargement of the eclipse phases in the LC of T-Lyr1-17236,
as recorded by the IAC80 (black) and the HATNet (grey) telescopes ($I$-band).
The solid line shows the best-fit JKT-EBOP model, for which the residuals
are displayed at the bottom.}
\label{figEclipse_I}
\end{figure}

\begin{figure}
\plotone{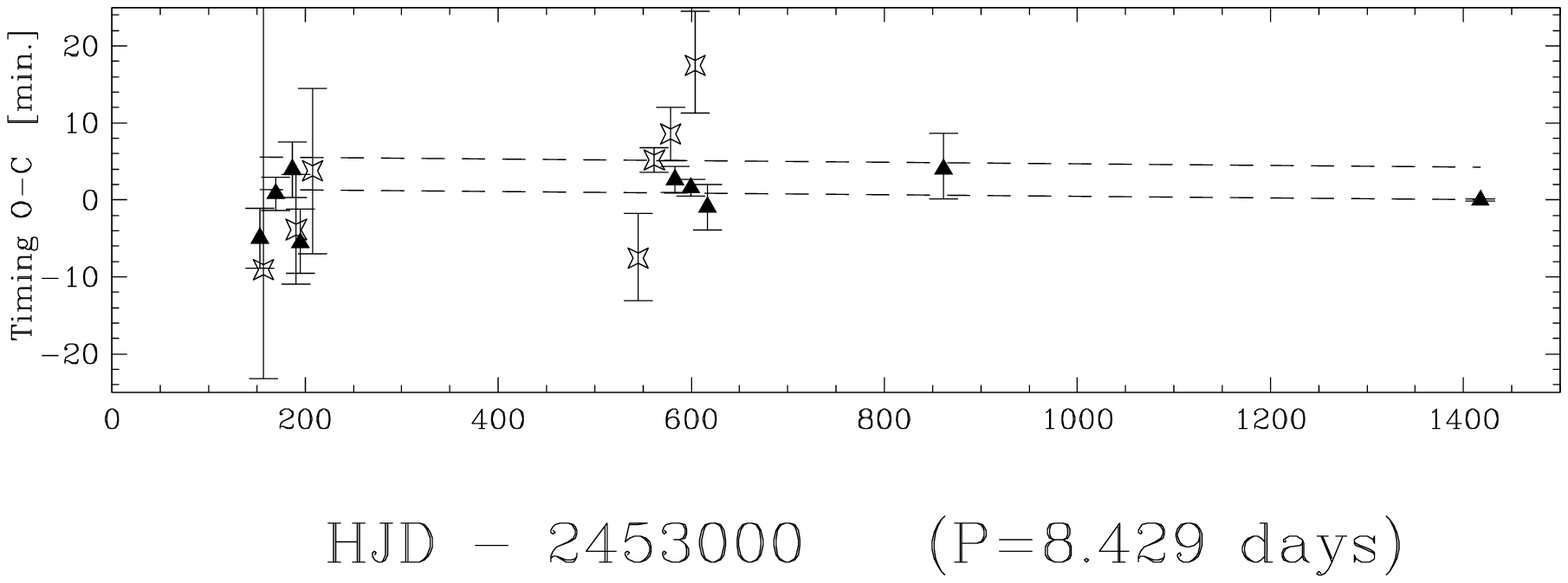} \caption{Eclipse timing ($O\!-\!C$) measurements
of T-Lyr1-17236.  The solid triangles indicate primary eclipses,
and the starred symbols indicate secondary eclipses. The large
error bars are generally due to eclipses that are constrained by
only a few observations, or for which only the ingress or egress
was observed.  The cluster of points at the very left (HJD $<
2,\!453,\!300$) are measurements from HATNet, the single data
point at HJD $2,\!454,\!417$ is from the IAC80, and the remaining
data are from Sleuth and PSST. The two parallel dashed lines
indicate the expected $O\!-\!C$ location of the primary (bottom)
and secondary (top) eclipses, in the best-fit eccentric model
($|e\cos\omega| \simeq 0.0005$). This eccentric model provides
only a very small improvement in the fit compared to the circular
model (F-test: $\chi^2_{\nu,circ} / \chi^2_{\nu,ecc} \simeq 1.29$,
indicating a $p \simeq 0.33$ significance).} \label{figTiming}
\end{figure}

\begin{figure}
\plotone{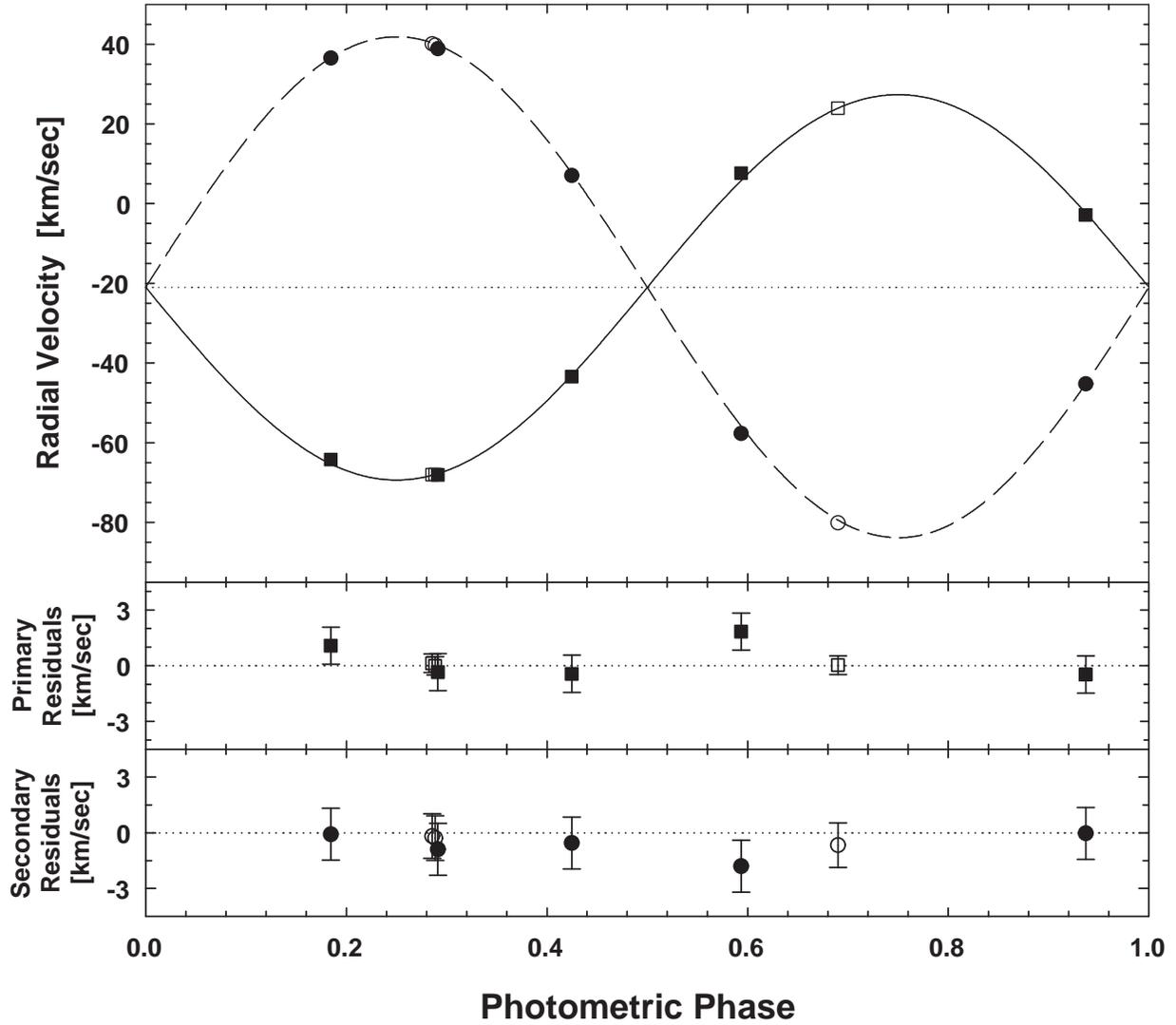}
\caption{Radial velocity measurements of
T-Lyr1-17236, relative to GJ~182, shown as a function of orbital
phase. The velocities of the primary component are represented
with squares, and those of the secondary with circles. The filled
symbols correspond to data taken with NIRSPEC, and the open
symbols represent TRES measurements.  Residuals from the model fit
are shown below for the primary and secondary components.}
\label{figPhasedRV}
\end{figure}

\begin{figure}
\epsscale{0.8}
\plotone{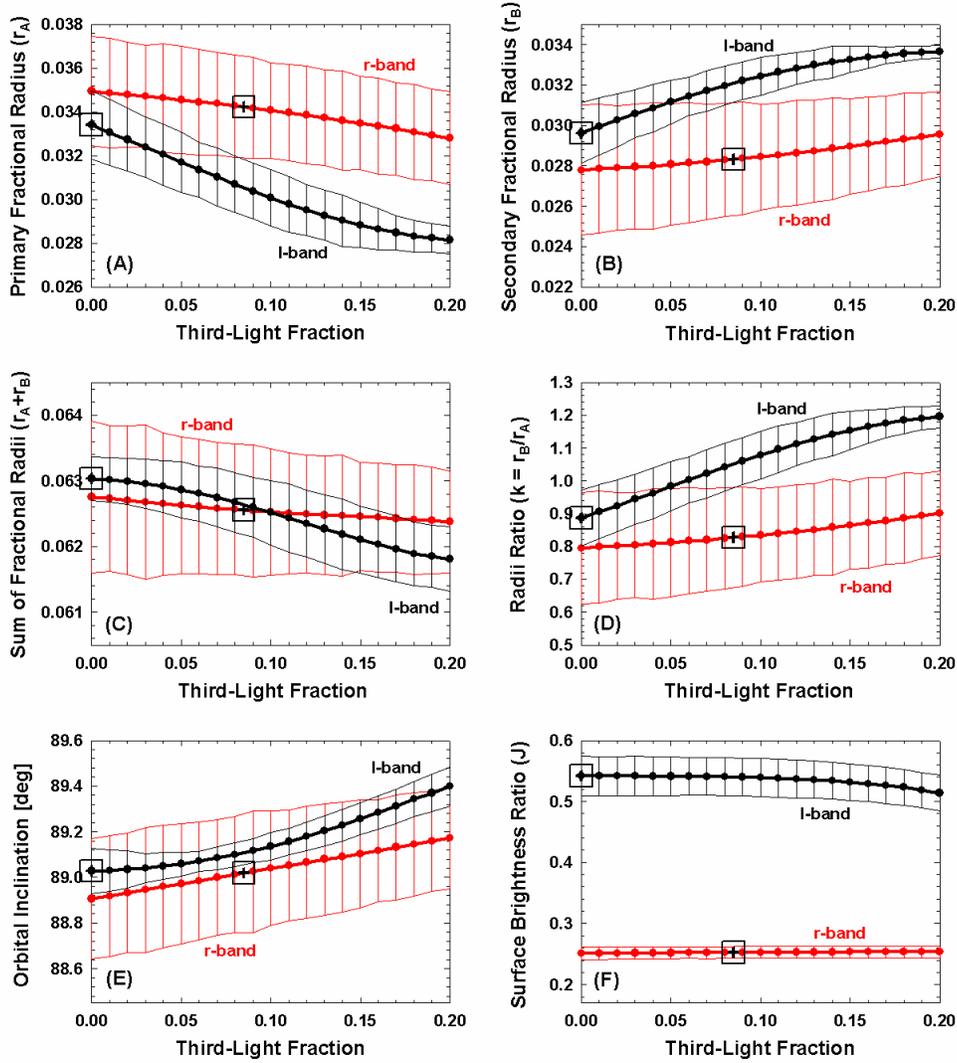}
\caption{The JKT-EBOP parameter fits over a range of values for the
third light fraction. The panels show the best-fit values and
uncertainties for: (A) The fractional radii of the primary ($r_A$),
and (B) secondary ($r_B$) components; (C) The sum of the fractional
radii ($r_A + r_B$), and (D) the radius ratio ($k = r_B / r_A$); (E)
The binary orbital inclination ($i$), and (F) the central surface
brightness ratio ($J$, secondary over primary). Note that in contrast
to other panels, panel (F) shows distinct values for the $I$- and
$r$-band LCs. This is expected since the two components have
different colors, and therefore different relative fluxes through
different filters. In all cases the estimated third light fractions
for the $r$-band and the $I$-band LCs are indicated by boxes.}
\label{fig3light}
\end{figure}

\begin{figure}
\plotone{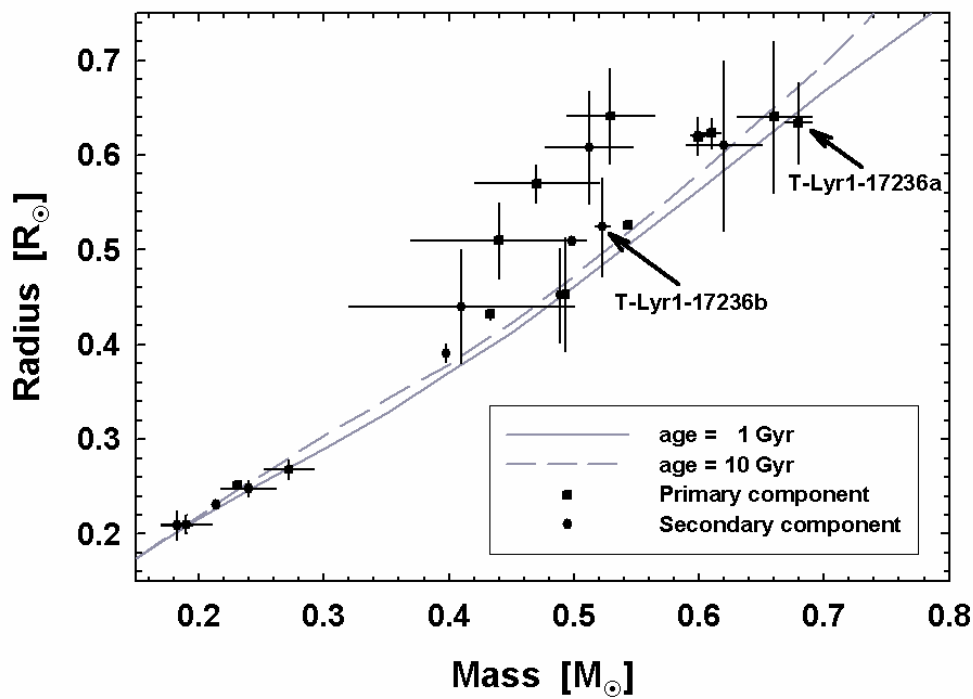}
\caption{Mass-radius diagram for T-Lyr1-17236 and other low-mass
eclipsing binaries under 0.8~$M_{\sun}$ from
Table~\ref{tablePreviousEBs}.  Theoretical isochrones for solar
metallicity from \cite{Baraffe98} are shown for ages of 1 and 10 Gyr.
The components of T-Lyr1-17236 are indicated with arrows. Most of
these binary components (particularly those with smaller
uncertainties) display a systematic offset in which their measured
radii are larger than predicted from models.}
\label{figIsochrones}
\end{figure}

Finally, we note that our findings in this paper confirm the
accuracy of the MECI algorithm (see Figure~\ref{figMECI}), which
can be further used to find additional long-period low-mass EBs,
and indeed a variety of other interesting targets. We have shown
in a recent paper \citep{Devor08} how this can be done with
comparable ease by systematically searching the ever-growing body
of LC survey datasets. We hope that this new approach for locating
rare EBs will motivate additional studies of these vast, largely
untapped datasets, which likely harbor a wealth of information on
the formation, structure, dynamics, and evolution of stars.

\acknowledgments

We would like to thank Joel Hartman and Doug Mink for their help
in operating a few of the software analysis tools used for this
paper, and we would like to thank Sarah Dykstra for her editorial
assistance. Valeri Hambaryan provided expert assistance in
examining archival ROSAT images of T-Lyr1-17236, for which we are
grateful, and we thank the referee for a number of helpful
comments that have improved the paper. GT acknowledges partial
support from NSF grant AST-0708229 and NASA's MASSIF SIM Key
Project (BLF57-04). This research has made use of NASA's
Astrophysics Data System Bibliographic Services, as well as the
SIMBAD database operated at CDS, Strasbourg. This publication also
used data products from the Two Micron All Sky Survey, which is a
joint project of the University of Massachusetts and the Infrared
Processing and Analysis Center/California Institute of Technology,
and is funded by NASA and NSF. Some of the data presented herein
were obtained at the W.M. Keck Observatory, which is operated as a
scientific partnership among Caltech, the University of California
and NASA. The Observatory was made possible by the generous
financial support of the W.M. Keck Foundation. The authors wish to
recognize and acknowledge the very significant cultural role and
reverence that the summit of Mauna Kea has always had within the
indigenous Hawaiian community. We are most fortunate to have the
opportunity to conduct observations from this mountain.

{}

\clearpage

\begin{deluxetable}{lcl}
\tabletypesize{\tiny}
\tablecaption{Periods of well characterized main sequence EBs with both component masses below 0.8 $M_{\sun}$}
\tablewidth{0pt}
\tablehead{\colhead{Name} & \colhead{Period [days]} & \colhead{Citation}}
\startdata
OGLE BW5 V38\tablenotemark{a} & 0.198 & \citet{Maceroni04}\\
RR Caeli\tablenotemark{b}     & 0.304 & \citet{Maxted07}\\
NSVS01031772 & 0.368 & \citet{LopezMorales06}\\
SDSS-MEB-1   & 0.407 & \citet{Blake07}\\
GU~Boo       & 0.489 & \citet{LopezMorales05}\\
2MASS J04463285+1901432 & 0.619 & \citet{Hebb06}\\
YY~Gem       & 0.814 & \citet{Kron52, Torres02}\\
T-Her0-07621 & 1.121 & \citet{Creevey05}\\
CM~Dra       & 1.268 & \citet{Lacy77CM, Metcalfe96}\\
UNSW-TR-2    & 2.117 & \citet{Young06}\\
2MASS J01542930+0053266 & 2.639 & \citet{Becker08}\\
CU~Cnc       & 2.771 & \citet{Delfosse99, Ribas03}\\
\enddata
\label{tablePreviousEBs}
\tablenotetext{a}{This binary might not be detached, as its components
seem to be undergoing significant mutual heating and tidal interactions due
to their proximity ($a=1.355 \pm 0.066 R_{\sun}$).}
\tablenotetext{b}{This is an unusual case of an EB containing a
white-dwarf (primary) and an M-dwarf (secondary). As such, the
primary component is likely to have transferred mass to the
secondary component, and perhaps even enveloped it during the
red-giant phase of its evolution.}
\end{deluxetable}

\begin{deluxetable}{llc}
\tabletypesize{\tiny}
\tablecaption{Catalog information for T-Lyr1-17236}
\tablewidth{0pt}
\tablehead{\colhead{Source Catalog} & \colhead{Parameter} & \colhead{Value}}
\startdata
2MASS\tablenotemark{a}\ \ \ \ \ & $\alpha$ (J2000)      & 19:07:16.621\\
2MASS & $\delta$ (J2000)      & +46:39:53.21\\
USNO-B\tablenotemark{b} & $B$ mag & 16.11 $\pm$ 0.2\\
GSC2.3\tablenotemark{c} & $V$ mag & 14.37 $\pm$ 0.28\\
USNO-B & $R$ mag                  & 14.41 $\pm$ 0.2\\
CMC14\tablenotemark{d}  & $r'$ mag & 14.073 $\pm$ 0.029\\
2MASS  & $J$ mag  & 12.019 $\pm$ 0.015\\
2MASS & $H$ mag         & 11.399 $\pm$ 0.015\\
2MASS & $K_s$ mag       & 11.235 $\pm$ 0.015\\
USNO-B & $\mu_\alpha$ (${\rm mas\,yr^{-1}}$) &  $-$2 $\pm$ 3\\
USNO-B & $\mu_\delta$ (${\rm mas\,yr^{-1}}$) & $-$28 $\pm$ 2\\
2MASS & identification    & 19071662+4639532\\
CMC14 & identification    & 190716.6+463953\\
GSC2.3 & identification   & N2EH033540\\
USNO-B & identification   & 1366-0314305
\enddata
\label{tableObjectParam}
\tablenotetext{a}{Two Micron All Sky Survey catalog \citep{Skrutskie06}.}
\tablenotetext{b}{U.S. Naval Observatory photographic sky survey \citep{Monet03}.}
\tablenotetext{c}{Guide Star Catalog, version 2.3.2 \citep{Morrison01}.}
\tablenotetext{d}{Carlsberg Meridian Catalog 14 \citep{Evans02}.}
\end{deluxetable}

\begin{deluxetable}{ccccll}
\tabletypesize{\tiny}
\tablecaption{Radial velocity measurements for T-Lyr1-17236 in the barycentric frame, relative to GJ~182}
\tablewidth{0pt}
\tablehead{\colhead{Epoch (BJD)} &
\colhead{\begin{tabular}{c} Primary RV\\ (\kms) \end{tabular}} &
\colhead{\begin{tabular}{c} Secondary RV\\ (\kms) \end{tabular}} &
\colhead{\begin{tabular}{c} Exposure Time \\ (sec) \end{tabular}} &
\colhead{Template} &
\colhead{Instrument}}
\startdata
2453927.9400   &  \phn$-$2.87  &  $-$45.24  &  480 & GJ 182  &  NIRSPEC\\
2453930.9258   &  $-$68.09  &  \phs38.85    &  900 & GJ 182  &  NIRSPEC\\
2453946.8846   &  $-$64.26  &  \phs36.53    &  600 & GJ 182  &  NIRSPEC\\
2453948.9100   &  $-$43.45  &  \phs\phn7.03    &  420 & GJ 182  &  NIRSPEC\\
2454312.7985   &  \phs\phn7.66    &  $-$57.68  &  480 & GJ 182  &  NIRSPEC\\
2454372.6179   &  \phs23.99    &  $-$80.14  &  900 & GJ 15A  &  TRES\\
2454377.6382   &  $-$68.03  &  \phs40.10    & 1000 & GJ 15A  &  TRES\\
2454377.6624   &  $-$67.97  &  \phs39.73    & 1000 & GJ 15A  &  TRES
\enddata
\label{tableLyr17236RVs}
\end{deluxetable}

\begin{deluxetable}{cccc}
\tabletypesize{\tiny}
\tablecaption{Eclipse timings measured for T-Lyr1-17236}
\tablewidth{0pt}
\tablehead{\colhead{Eclipse Type} & \colhead{Epoch (HJD)} & \colhead{O-C [sec]} & \colhead{Data Source}}
\startdata
Primary    & 2453152.96121 &   $-299^{+232}_{-236}$  & HATNet\\
Secondary  & 2453157.17593 &   $-546^{+6868}_{-849}$ & HATNet\\
Primary    & 2453169.82009 &   $48^{+126}_{-131}$ & HATNet\\
Primary    & 2453186.67897 &   $237^{+214}_{-221}$ & HATNet\\
Secondary  & 2453190.89369 &   $-231^{+431}_{-423}$ & HATNet\\
Primary    & 2453195.10841 &   $-333^{+263}_{-238}$ & HATNet\\
Secondary  & 2453207.75258 &   $225^{+642}_{-648}$ & HATNet\\
Secondary  & 2453544.93022 &   $-452^{+346}_{-332}$ & Sleuth \\
Secondary  & 2453561.78910 &   $312^{+97}_{-98}$ & Sleuth + PSST\\
Secondary  & 2453578.64798 &   $515^{+206}_{-208}$ & Sleuth\\
Primary    & 2453582.86270 &   $159^{+99}_{-98}$ & Sleuth\\
Primary    & 2453599.72158 &   $94^{+64}_{-64}$ & Sleuth\\
Secondary  & 2453603.93630 &   $1047^{+424}_{-371}$ & Sleuth + PSST\\
Primary    & 2453616.58046 &   $-57^{+175}_{-175}$ & Sleuth\\
Primary    & 2453861.03425 &   $238^{+280}_{-233}$ & PSST\\
Primary    & 2454417.37736 &   $-1^{+10}_{-10}$ & IAC80
\enddata
\label{tableLyr17236Timing}
\end{deluxetable}

\begin{deluxetable}{lcc}
\tabletypesize{\tiny}
\tablecaption{Photometric parameters of T-Lyr1-17236}
\tablewidth{0pt}
\tablehead{\colhead{Parameter} & \colhead{Symbol} & \colhead{Value}}
\startdata
Period (days)               & $P$   & 8.429441 $\pm$ 0.000033\\
Epoch of eclipse (HJD)      & $t_0$ & 2453700.87725 $\pm$ 0.00041\\
Primary fractional radius   & $r_A$ & 0.0342 $\pm$ 0.0023\\
Secondary fractional radius & $r_B$ & 0.0283 $\pm$ 0.0028\\
Orbital inclination [deg]   & $i$   & 89.02 $\pm$ 0.26\\
Eccentricity                & $e$   & 0.0 (fixed)\\
Sum of fractional radii     & $r_A+r_B$ & 0.06256 $\pm$ 0.00095\\
Ratio of radii ($R_B/R_A$)  & $k$   & 0.83 $\pm$ 0.15\\
Light ratio ($r$-band)     & $L_B/L_A$ &   0.173 $\pm$ 0.073\\
Surface brightness ratio ($r$-band) & $J_B/J_A$ & 0.2525 $\pm$ 0.0099
\enddata
\label{tablePhotoParam}
\end{deluxetable}

\begin{deluxetable}{lcc}
\tabletypesize{\tiny}
\tablecaption{Spectroscopic parameters of T-Lyr1-17236}
\tablewidth{0pt}
\tablehead{\colhead{Parameter} & \colhead{Symbol} & \colhead{Value}}
\startdata
Primary radial velocity semi-amplitude ($\kms$)   & $K_A$          & 48.36 $\pm$ 0.23\phn\\
Secondary radial velocity semi-amplitude ($\kms$) & $K_B$          & 62.86 $\pm$ 0.46\phn\\
Barycentric radial velocity, relative to GJ~182\tablenotemark{a}\ \ ($\kms$) & $V_\gamma$  & $-$21.01 $\pm$ 0.18\phs\\
Binary separation with projection factor ($R_{\sun}$)      & $a \sin i$     & 18.526 $\pm$ 0.083\phn\\
Primary mass with projection factor ($M_{\sun}$)           & $M_A \sin^3 i$ & 0.6792 $\pm$ 0.0107\\
Secondary mass with projection factor ($M_{\sun}$)         & $M_B \sin^3 i$ & 0.5224 $\pm$ 0.0061\\
Mass ratio ($M_B/M_A$)                                     & $q$            & 0.7692 $\pm$ 0.0069
\enddata
\label{tableSpecParam}
\tablenotetext{a}{\citet{Montes01} list the radial velocity of GJ~182 as $+32.4 \pm 1.0$~\kms.}
\end{deluxetable}

\begin{deluxetable}{lccc}
\tabletypesize{\tiny}
\tablecaption{System parameters of T-Lyr1-17236}
\tablewidth{0pt}
\tablehead{\colhead{Parameter} & \colhead{Symbol} & \colhead{Component A} & \colhead{Component B}}
\startdata
Mass ($M_{\sun}$)          & $M$     & 0.6795 $\pm$ 0.0107 & 0.5226 $\pm$ 0.0061\\
Radius ($R_{\sun}$)        & $R$     & 0.634 $\pm$ 0.043   & 0.525 $\pm$ 0.052\\
Log surface gravity (cgs)  & $\log g$& 4.666 $\pm$ 0.059   & 4.718 $\pm$ 0.086\\
Semimajor axis ($10^6$ km) & $a$     & 5.606 $\pm$ 0.027   & 7.288 $\pm$ 0.053 \\
Maximum rotational velocity\tablenotemark{a}\ \ (\kms) & $v \sin i_r$ & $5.6 \pm 2.0$ & $5.1 \pm 2.3$ \\
Synchronized rotational velocity\tablenotemark{a}\ \ (\kms) & $(v \sin i_r)_{\rm sync}$ & $3.81 \pm 0.26$ & $3.15 \pm 0.31$ \\
Absolute visual magnitude\tablenotemark{b}\ \ (mag) & $M_V$ & 8.03 & 9.67 \\
Bolometric luminosity\tablenotemark{b}\ \ ($L_{\sun}$) & $L$ & 0.110 & 0.039 \\
Effective temperature\tablenotemark{b}\ \ (K)  & $T_{\rm eff}$ & 4150 & 3700 \\
Distance\tablenotemark{b}\ \ (pc)             & $D$ & \multicolumn{2}{c}{230 $\pm$ 20}
\enddata
\tablenotetext{a}{See description in \S\ref{secPhysical}.}
\tablenotetext{b}{Inferred using stellar evolution models by
\cite{Baraffe98} assuming solar metallicity and an age of
$2.5$~Gyr.}
\label{tableSystemParam}
\end{deluxetable}

\newpage

\end{document}